# Fano resonance control in a photonic crystal structure and its application to ultrafast switching

Yi Yu,[†] Mikkel Heuck, Hao Hu, Weiqi Xue, Christophe Peucheret,
Yaohui Chen, Leif Katsuo Oxenløwe, Kresten Yvind, and Jesper Mørk

*DTU Fotonik, Department of Photonics Engineering, Technical University of Denmark, DK-2800 Kongens Lyngby, Denmark*

[†]E-mail address: yiyu@fotonik.dtu.dk

**Fano resonances appear in quantum mechanical as well as classical systems as a result of the interference between two paths: one involving a discrete resonance and the other a continuum. Compared to a conventional resonance, characterized by a Lorentzian spectral response, the characteristic asymmetric and "sharp" spectral response of a Fano resonance is suggested to enable photonic switches and sensors with superior characteristics. While experimental demonstrations of the appearance of Fano resonances have been made in both plasmonic and photonic-crystal structures, the control of these resonances is experimentally challenging, often involving the coupling of near-resonant cavities. Here, we experimentally demonstrate two simple structures that allow surprisingly robust control of the Fano spectrum. One structure relies on controlling the amplitude of one of the paths and the other uses symmetry breaking. Short-pulse dynamic measurements show that besides drastically increasing the switching contrast, the transmission dynamics itself is strongly affected by the nature of the resonance. The influence of slow-recovery tails implied by a long carrier lifetime can thus be reduced using a Fano resonance due to a hitherto unrecognized reshaping effect of the nonlinear Fano transfer function. For the first time, we present a system application of a Fano structure, demonstrating its advantages by the experimental realization of 10 Gbit/s all-optical modulation with bit-error-ratios on the order of $10^{-7}$ for input powers less than 1 mW. These results represent a significant improvement compared to the use of a conventional Lorentzian resonance.**

Ultra-compact photonic structures that perform optical signal processing such as modulation and switching at high-speed with low-energy consumption are essential for enabling integrated photonic chips that can meet the growing demand for information capacity[1]. The photonic-crystal (PhC) platform is promising in this respect, since it allows the integration of lasers[2], modulators[3], switches[4] and detectors[5]. It remains, however, an important task to identify and demonstrate PhC structures that can meet the low-energy and high-bandwidth requirement. In cavity-based switches an applied control signal changes the refractive index of the cavity, thereby shifting the cavity resonance and modulating the transmission of the data signal. The shape of the transmission spectrum is then very important, since it determines the refractive index shift, and thereby the control energy required for achieving a certain on-to-off ratio. The long tails of a Lorentzian spectrum thus imply large switching energies as well as slow dynamic components dominated by a long carrier lifetime[6], which prevent high-bit-rate signal-processing unless the slow dynamics can be efficiently suppressed[7]. In contrast, a Fano resonance[8] has an asymmetric spectrum, featuring a large transmission change within a narrow wavelength range determined by the transition from constructive to destructive interference between the discrete resonance and the continuum, thus enabling low-energy switching[9].

While Fano effects, recognized as a universal phenomenon[8-15], have been investigated in various nanostructures[9-21], most works culminate in the demonstration of an asymmetric transmission spectrum. In Ref. 16, very sharp Fano resonance were thus achieved. However, like many previous works[9, 17, 18], the structure is not compact and does not have an in-line configuration, making it unsuitable for on-chip signal-processing. Additionally, the Fano resonance is highly sensitive to fabrication and experimental conditions, which requires very precise control of parameters such as the incident angle of the light beam and the displacement between the PhC layers.

In Ref. 19, a multi-cavity PhC structure was proposed and the asymmetric line shape was realized through an induced absorption effect. Compared to a Lorentzian structure, larger switching contrast was achieved, as demonstrated by pump-probe measurements. However, this multi-cavity scheme, where the asymmetry is characterized by more than two parameters, requires the control of the resonance frequencies of two cavities as well as their mutual coupling [20, 21], making it sensitive to the fabrication process. No system experiments employing Fano resonances were yet, to the best of our knowledge, reported, mainly due to the fact that the control of these resonances is experimentally challenging.

Here, we experimentally demonstrate a simple PhC nanocavity structure that allows easy and surprisingly robust control of the Fano spectrum. Through short-pulse dynamic measurements, we demonstrate for the first time that besides drastically increasing the modulation contrast, the slow-recovery tails of the modulated signal are suppressed when employing a Fano resonance. The advantages of the Fano structure are further verified by applying the device for 10 Gbit/s all-optical modulation, which is the first application of a Fano structure on the system level, achieving bit-error-ratios (BER) on the order of $10^{-7}$ for input powers less than 1 mW, which represents an improvement by several orders of magnitude compared to the use of a conventional Lorentzian resonance[22].

The investigated structure is illustrated in Fig. 1a and consists of a PhC line-defect waveguide coupled to a point-defect nanocavity. In contrast to earlier work[23], we add below the nanocavity a partially transmitting element (PTE) that allows controlling the amplitude of the continuum-path. This concept was outlined theoretically[24, 25], but not yet realized experimentally.

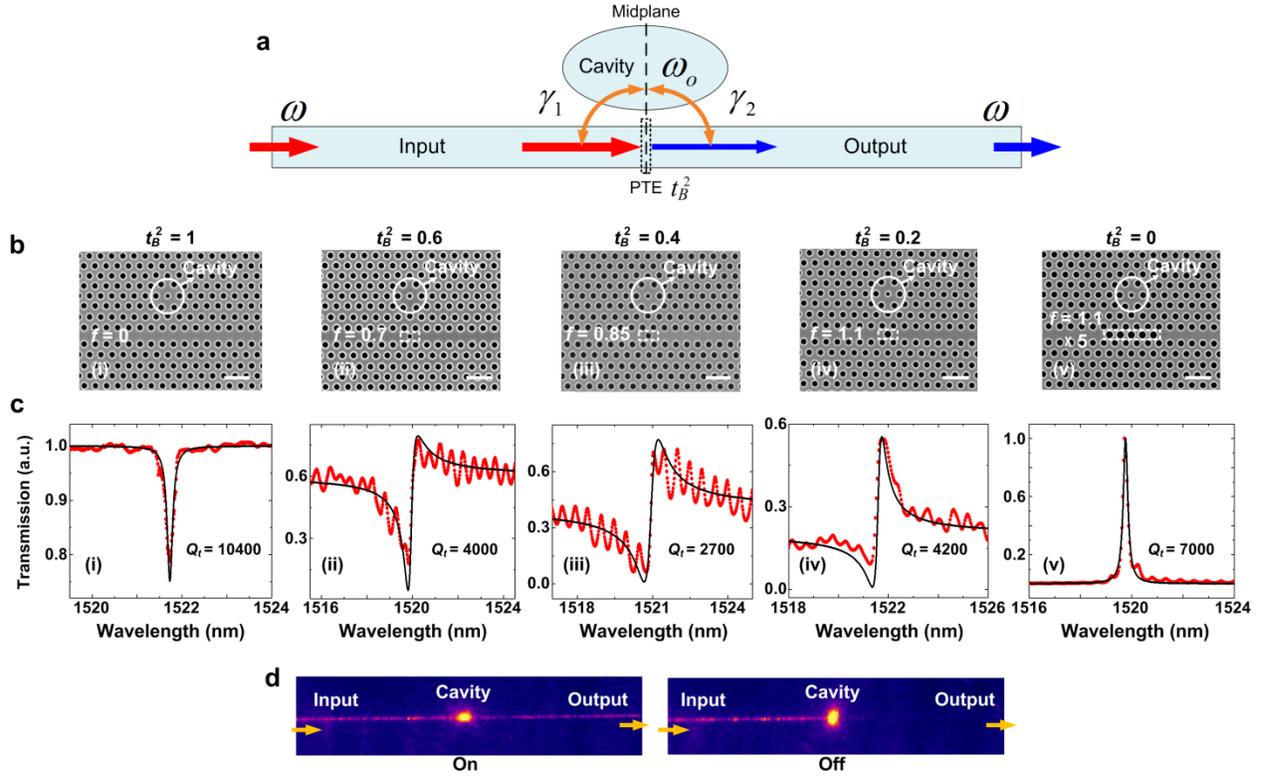

**Figure 1 | Cavity-waveguide system with mirror symmetry. a** Transmission of a signal at frequency $\omega$ through a waveguide that is side-coupled to a cavity with resonance frequency $\omega_0$. The field in the cavity couples to the in- and output ports with decay rates $\gamma_1$ and $\gamma_2$. The structure is symmetric around the mid-plane (vertical dashed line). A partially transmitting element (PTE, dashed rectangle) with power transmission $t_B^2$ is placed at the mid-plane. **b** SEM images of fabricated PhC structures with lattice constant $a$ = 447 nm, air-hole radius $R$ = 116±0.8 nm, and membrane thickness $h$ = 340 nm. The scale bar corresponds to 1 μm. The small air-hole in the cavity has radius $R_c$ = 0.85±0.01$R$. The PTE is implemented by etching one or several blockade-hole (s) (BH) with a radius of $R_B$ in the waveguide, and $t_B^2$ is tuned by varying the BH number or its radius, quantified by the ratio $f = R_B/R$. **c** Measured (red dots) and theoretical (black lines) transmission spectra, normalized to one at the transmission maximum for the Lorentzian resonance ($t_B^2 = 0$). From the inverse Lorentzian shape ($t_B^2 = 1$) and the three-dimensional finite difference time domain (FDTD) simulations, the cavity exhibits an intrinsic quality-factor $Q_v$ of ~1.2×10$^4$ and a small mode volume of 0.44($\lambda/n$)$^3$. **d** Infrared camera images taken at "on" and "off" states of the Fano resonance.

We realized the proposed structure in an InP air-embedded PhC membrane, with the transmission coefficient $t_B^2$ of the PTE determined by the size of the air-hole(s) in the waveguide (see Supplementary Information (SI), section A.2). Fig. 1b shows scanning electron microscope (SEM) images of a set of structures and Fig. 1c shows corresponding experimental transmission spectra. By changing $t_B^2$ from the value of one (open waveguide) to zero (blocked waveguide), we observe the smooth development of the spectrum from a symmetric inverse Lorentzian shape (reflecting destructive interference between the cavity and the waveguide path at resonance), via characteristic asymmetric Fano shapes (reflecting destructive and constructive interference on opposite sides of the resonance), to the well-known Lorentzian profile (reflecting that the only path from input to output is via the cavity).

Fig. 1c also shows theoretical results obtained using temporal coupled mode theory[24-27], allowing the transmission to be expressed as (SI, section A.1):

$$t(\omega) = \frac{t_B(\omega_0 - \omega) \pm \sqrt{4\gamma_1\gamma_2 - t_B^2(\gamma_1 + \gamma_2)^2} - jt_B\gamma_v}{j(\omega_0 - \omega) + \gamma_t}. \quad (1)$$

The intrinsic and total loss rates, $\gamma_v$ and $\gamma_t = \gamma_1 + \gamma_2 + \gamma_v$, can be related to the cavity intrinsic and total quality-factors, $Q_v$ and $Q_t$, as $\gamma_v = \omega_o/2Q_v$ and $\gamma_t = \omega_o/2Q_t$. If the system has mirror symmetry (Fig. 1a), $\gamma_1 = \gamma_2 = \gamma_{in}/2$ and Eq. (1) becomes:

$$t(\omega) = t_B^2 \pm jt_B r_B - \frac{\gamma_{in}}{j(\omega_o - \omega) + \gamma_t}, \quad (2)$$

where $r_B = \sqrt{1 - t_B^2}$. The positive (negative) sign in Eq. (2) corresponds to a cavity mode that is odd (even) with respect to the mid-plane. We define the parity of a Fano line by whether the transmission minimum is red- or blue shifted relative to the maximum. For the odd cavity mode used in the experiment we thus always observe blue parity. The good agreement between experiment and theory confirms that we observe a Fano resonance and that its features can be simply controlled via the PTE transmission. The value of $Q_t$, given in Fig. 1c, varies with blockade hole (BH) size due to the modified scattering process. The oscillations in the measured spectra with a period of 0.5-0.6 nm are Fabry-Perot fringes due to reflections at the waveguide end-facet and the cavity. We characterized a number of devices and found that the Fano shape is reproducible and readily controlled via the air-hole, demonstrating the robustness of the suggested structure.

Breaking the symmetry of the structure is another way of controlling the Fano line shape. Thus, by displacing the PTE within a few lattice constants, Figs. 2a, 2b, the measured transmission spectra show that it is possible to realize Lorentzian (Fig. 2c(ii)) as well as Fano line shapes (Figs. 2c(i,iii,iv)). While the parity of the Fano line is determined by the cavity mode symmetry in the symmetric case, cf. Fig. 1, the displacement of the PTE allows independent control over the parity, as seen by comparing e.g. Figs. 2c(iii) and 2c(iv). Fig. 2c also shows theoretical transmission spectra obtained using Eq. (1). Due to the asymmetry, the coupling coefficients $\gamma_1$ and $\gamma_2$ now differ and the decay ratio $\gamma_2/\gamma_1$ varies with $D$, giving rise to the change of the line shape. The variation of the decay rates can be interpreted as a result of interference (SI, Section A.1). It was recently shown theoretically[27] that the decay ratio $\gamma_2/\gamma_1$ is bounded from below by $(1-r_B)/(1+r_B)$ and from above by $(1+r_B)/(1-r_B)$. Interestingly, we find that when $D = 0.5a$ (Fig. 2b(ii)), both $\gamma_2/\gamma_1$ and $t_B^2$ approach unity even when the structure is asymmetric, resulting in the transmission exhibiting an almost symmetric Lorentzian-dip with a $Q_t$ very different from the case in Fig. 1c(i). When $t_B^2 = 0$ (Fig. 2c(v)), the line shape is Lorentzian independently of the PTE position, since light can only pass through the cavity but not the waveguide, quenching the interference effects responsible for asymmetric line shapes.

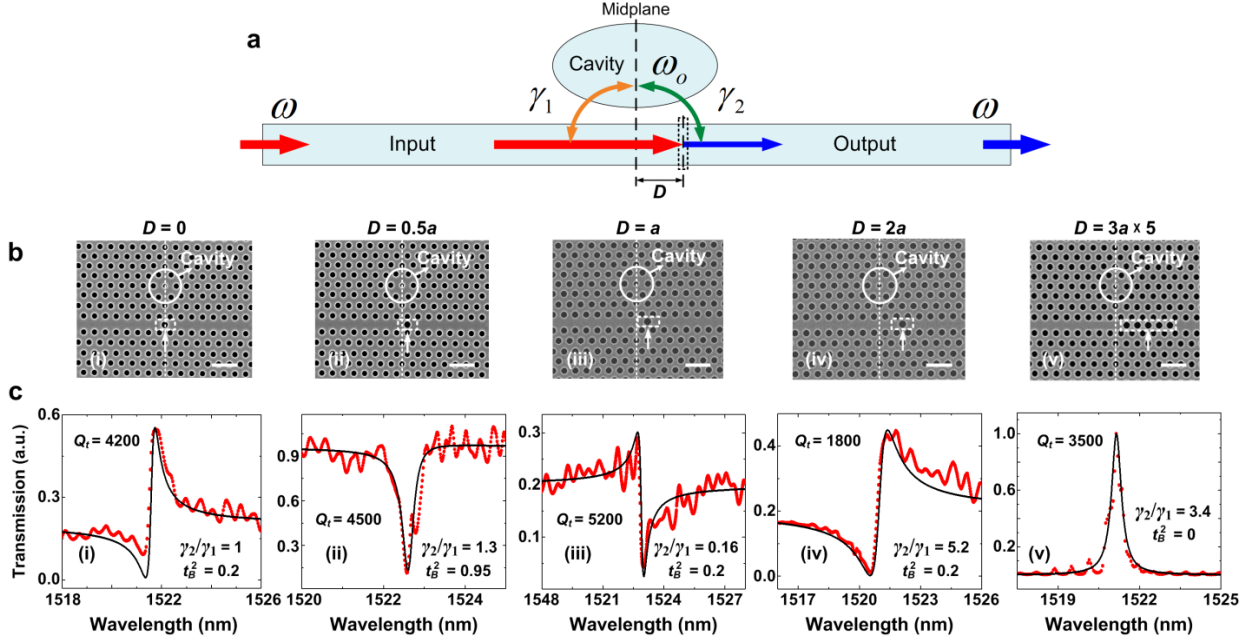

**Figure 2 | Cavity-waveguide system with broken symmetry. a** Similar to Fig. 1a, but the PTE is offset by a distance $D$ (in units of the lattice constant $a$) with respect to the midplane (vertical dashed line). **b** SEM images of the asymmetric PhC with fixed BH radius ($f = 1.1$) and varying $D$. The scale bar corresponds to 1 μm. **c** Corresponding measured (red dots) and theoretical (black lines) transmission spectra normalized to one at the transmission maximum for the Lorentzian resonance. When $D$ is varied, the values of $t_B^2$ and $\gamma_2 / \gamma_1$ change depending on the position of the PTE relative to the electromagnetic field pattern.

In general, when a signal is transmitted through a switch (or modulator), not only the intensity but also the phase of the signal is modified, as quantified by the chirp-parameter giving the ratio of phase and amplitude changes[21]. Fig. 3a presents what we believe to be the first measurement (see Methods) of the phase spectrum of Fano structures, showing how the spectrum changes depending on the PTE transmission. Fig. 3b shows the corresponding measured intensity spectra. For sufficiently low input power, the measured spectra agree very well with linear theory, where the resonance frequency is not perturbed by the energy build-up in the cavity. The results of Fig. 3 show that a small blueshift of the Fano resonance can lead to a larger intensity change than for a Lorentzian (compare the black curves in Figs. 3b(ii) and 3b(iii)), while the phase shifts are similar (compare the black curves in Figs. 3a(ii) and 3a(iii)). These results suggest that Fano structures are promising for realizing intensity modulators with large modulation contrast and small chirp-parameter (SI, section B.4).

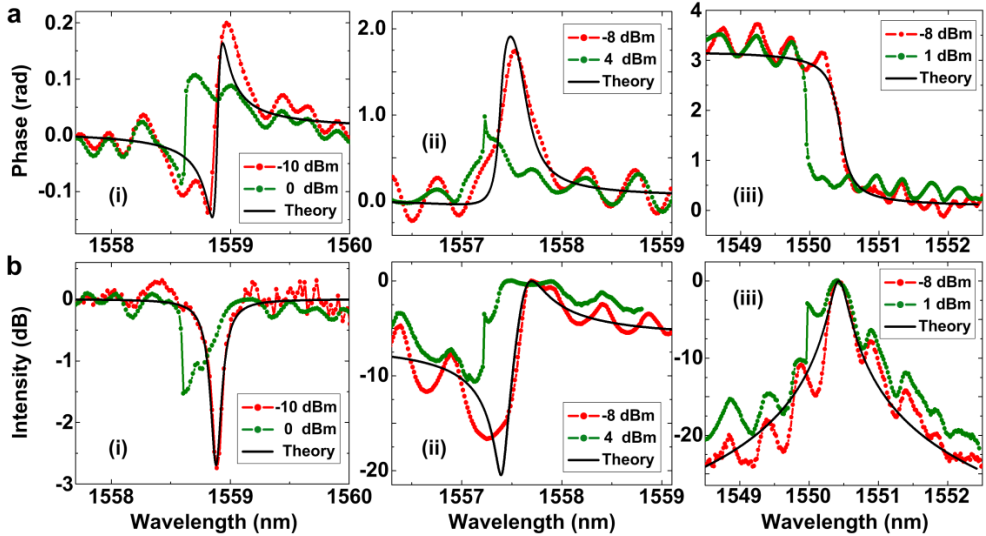

**Figure 3 | Linear and nonlinear phase and intensity spectra.** Measured **a** phase spectra and **b** intensity spectra of (i) open, (ii) partially-blocked and (iii) blocked structures for two different input powers. The intensity is normalized to one at its maximum value. Red and green dotted lines are experimental data at low and high input powers, respectively, and black solid lines represent theoretical fits to the low-power, linear case. The structures are symmetric and have total quality factors $Q_t$ of (i) $1.1\times10^4$ with $t_B^2 = 1$, (ii) $6.1\times10^3$ with $t_B^2 = 0.1$ and (iii) $8.6\times10^3$ with $t_B^2 = 0$, respectively. The three structures have similar $Q_v$ of $\sim 1.5\times10^4$.

For larger input power, see green curves in Fig. 3, the spectra change qualitatively due to an induced change of the resonance frequency. We show in SI, Section B.1, that the observed effects are consistent with a reduction of the refractive index due to plasma effects and band-filling caused by free carriers generated via two-photon absorption. The abrupt changes of output power and phase with wavelength occur due to a region of bistability located on the blue side of the resonance.

We also perform measurements of the nonlinear dynamics of a Fano ($\lambda_0 = 1556$ nm, $Q_t = 3.8\times10^3$, $t_B^2 = 0.2$) and a conventional Lorentzian structure ($\lambda_0 = 1540$ nm, $Q_t = 5.2\times10^3$, $t_B^2 = 0$), cf. Fig. 4a. A weak CW signal is injected into the device together with 12-ps pump pulses at a repetition rate of 5 GHz. We consider the switch-on case where the signal is blue-detuned with respect to the cold-cavity resonance. The color maps (left) show the temporal variation of the probe signal for different detunings, with the dashed lines corresponding to the specific traces (right), where the 90%-10% transmission decay time is shortest. The Fano structure shows a decay time of 30 ps with a switching contrast of 14 dB, as compared to 80 ps and 4 dB for the Lorentzian structure for the same pump pulse energy. It has previously been shown[4] that the carrier distribution generated by the pump in the nanocavity first relaxes on a very short time scale due to diffusion and subsequently has a slower recovery governed by surface and bulk recombination. This effect is important, since it is the slow dynamics that limit high-speed applications[6]. Comparing the transmission dynamics of the Lorentzian and the Fano structure, it is clearly seen that the slow tail is suppressed for the Fano structure.

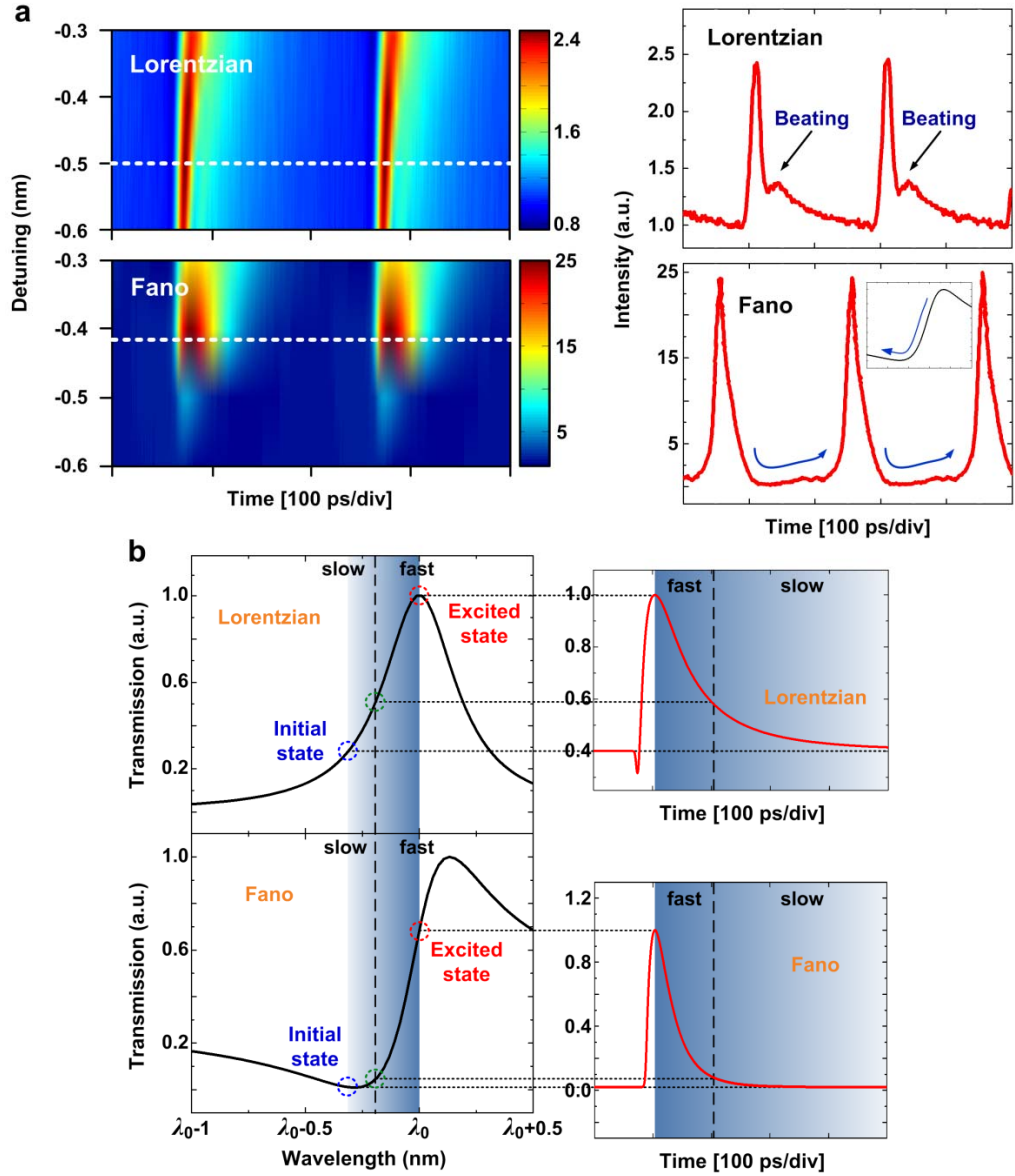

**Figure 4 | Dynamics of the Fano structure compared to the Lorentzian structure. a** Measured waveforms of the output signal for the Fano and Lorentzian structures. The pump pulses, with energy of ~110 fJ, are tuned to the cavity resonance and the signal is blue detuned by 0.3-0.6 nm. 2D color maps (left) show the time-resolved intensity (color scale) for different detunings. Time domain traces of the intensity are shown to the right for the detuning values (indicated by horizontal white dashed lines in the 2D maps) giving the shortest 90%-10% transmission decay time. The blue arrows shown for the Fano structure indicate the turning-point characteristics of the resonant profile shown in the inset. **b** Theoretical transmission (left) of Lorentzian ($t_B^2 = 0$) and Fano ($t_B^2 = 0.2$) structure versus wavelength detuning and corresponding transmission dynamics (right). The signal is first located at the initial state and then moves to the excited state due to pump excitation, after which it recovers back. The slow transmission recovery is suppressed in the Fano structure due to the small transmission variation around the initial state (= final state). The two structures have the same intrinsic and total quality-factors.

This effect can be explained as a consequence of the nonlinear transmission spectrum of the Fano structure, exhibiting vanishing slope at the transmission minimum, see Fig. 4b. A detailed analysis (SI, section B.3) based on temporal coupled mode theory further supports the conclusion that Fano structures are superior to Lorentzian structures for high-speed signal-processing. We notice that the Lorentzian structure exhibits an oscillatory characteristic in the initial decay, which is due to interference beating between the signal and cavity mode[6,29].

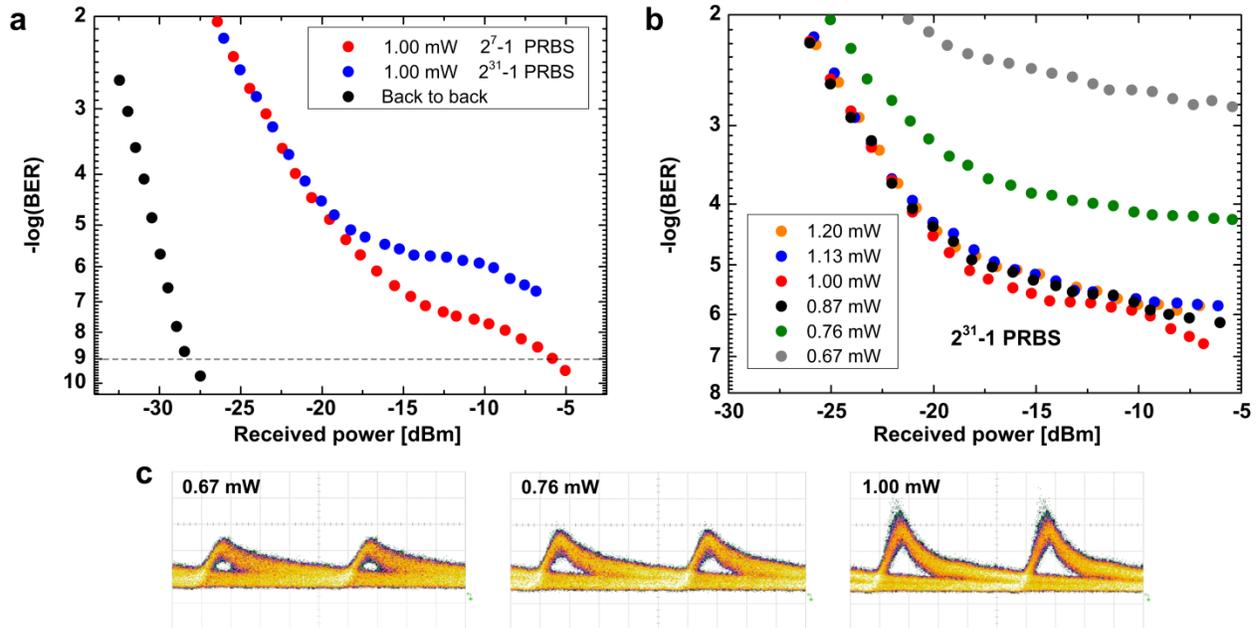

**Figure 5 | BER measurements for 10 Gbit/s all-optical switching using the Fano structure. a** BER measurements of the modulated signal (colored dots) and back-to-back signal (black dots). The 10 Gbit/s pump pulses with average power of 1 mW are modulated by PRBSs with sequence lengths of $2^7-1$ (red dots) and $2^{31}-1$ (blue dots), respectively. **b** BER measurements of the modulated signal for different average powers ranging from 0.67 to 1.2 mW with a PRBS of $2^{31}-1$. **c** Corresponding eye diagrams of the modulated signal with average power of 0.67, 0.76 and 1 mW.

Finally, to investigate the system properties, we characterize the BER when the Fano structure is used to modulate a weak continuous wave probe with a data-modulated pump signal. The pump pulses with a full width at half maximum of ~12 ps are on-off keying (RZ-OOK) modulated by a 10 Gbit/s pseudo-random binary sequence (PRBS) with the sequence lengths of $2^7-1$ and $2^{31}-1$. To alleviate the pump-probe cross-talk, we engineer the Fano resonance to achieve a relatively low $Q_t$ of $1.5 \times 10^3$ and small $t_B^2$ of 0.08 by tuning the width of the PhC waveguide (shifting the innermost two rows of air holes of the waveguide) and the BH size, which gives a relatively large spacing (~1.7 nm) between the resonance extreme, although this also reduces the spectral slope and thus increases the switching energy. The pump and signal are slightly red-detuned from the resonance maximum and minimum, respectively, to counteract a small mean thermal shift of the resonance.

Fig. 5a shows error free (BER<$10^{-9}$) all-optical switching for the 10 Gbit/s pump signal with a PRBS of $2^7-1$ and a coupled average power of 1 mW. For a longer PRBS sequence of $2^{31}-1$, the BER increases somewhat, but stays below $2 \times 10^{-7}$, well below the threshold when employing forward error correction (FEC) schemes. The larger BER for longer PRBS length and the error floor are mainly due to residual patterning effects. For comparison, using a device with a conventional Lorentzian resonance, the lowest BER values obtained were on the order of $10^{-3}$ even if a short PRBS of $2^7$-1 was used[22], thus clearly demonstrating the advantage of Fano structures for reducing the patterning effect. Note that the energy consumption can be significantly reduced if the cross-talk issue is addressed since the pump signal can then be placed closer to the resonant wavelength (located on the slope of the Fano line), drastically enhancing its coupling efficiency into the cavity. The relatively large power penalty originates from the residual patterning effect as well as a high coupling loss of the signal. Here the back-to-back BER curve is measured for the RZ-OOK pump signal with a small duty cycle (12%), which has relatively high receiver sensitivity. For a larger duty cycle of 33% or a non-return-to-zero OOK signal, which has lower receiver sensitivity, higher power level would therefore be needed for the back-to-back measurement and the power penalty due to the residual patterning effect would be smaller[30,31].

In Fig. 5b, the BER first decreases as the coupled input average power increases (from 0.67 to 1 mW), and then saturates (around 1 mW) followed by a small increase (from 1 to 1.2 mW). In all the cases, the measured BER stays below the FEC limit of $3.8 \times 10^{-3}$. The saturation phenomenon of the BER curve occurs since, for low pump energy, the switched signal quality is limited by the transmission noise due to a small switching contrast. For larger pump energy, besides the patterning effect, the BER is limited by pump leakage since the resonance drags part of the pump into the probe frequency band due to an enhanced adiabatic tuning[32]. This can be seen in the corresponding eye diagrams, Fig. 5c. An estimated extinction ratio of 5 (8.7) dB was obtained for an average power of 0.67 (1) mW. For 1 mW, more serious noise appears on the rising edges of the signal, representing enhanced cross-talk from the pump. These results demonstrate a very fundamental improvement enabled by the use of a Fano resonance as opposed to the traditional Lorentzian, which enables us, for the first time, to obtain such low BER values in such a simple all-optical switch using realistic pattern lengths. The PhC membrane structure offers many additional possibilities for improving the performance, e.g. the use of surface growth quantum wells and $SiO_2$ encapsulation[7] to reduce the carrier lifetime and improve the thermal conductivity, which would further suppress the slow recovery tail of the modulated signal and the associated patterning effects.

In conclusion, we have experimentally demonstrated a PhC waveguide-cavity structure that allows very robust control of the transmission line shape. Lorentzian and Fano line shapes are realized by varying the size of a single air-hole. Additional control of the parity of the Fano shape, independently of the symmetry of the cavity mode, was obtained by breaking the mirror symmetry of the structure. Very good agreement between experiment and theory was demonstrated, even extending to the phase response of the Fano structures, which was measured for the first time. Qualitatively different nonlinear and dynamical characteristics were observed for Lorentzian and Fano structures. In particular, the turning-point characteristic of Fano structures was demonstrated to allow the suppression of slow transmission dynamics, which limit the speed of conventional switches based on a Lorentzian response function. The Fano structure, enabled the first experimental demonstration of 10 Gbit/s RZ-OOK all-optical switching in a simple PhC cavity structure using realistic PRBS patterns ($2^{31}$-1), and achieving BER values as low as $2\times10^{-7}$, well below the FEC threshold with low energy consumptions (< 1 mW). The results thus demonstrate several advantages of switches employing Fano effects: much faster modulation speed, an order of magnitude larger modulation contrast and smaller chirp-parameter, making the suggested ultra-small Fano structure very promising for low-energy, ultra-fast on-chip signal-processing. The demonstrated breaking of mirror symmetry can be used to further reduce the mode volume since a blue parity Fano line, suitable for switching, can be realized using a smaller H0 cavity[33] containing only an even mode. Other improvements are possible, e.g. the combination with multi-mode[7,34] and multi-port[35] designs to relieve the pump-signal cross-talk limitation as well as employing quaternary InGaAsP material, where carriers are generated through a combination of two-photon and linear absorption[4], significantly reducing the energy consumption. Considering that the device properties are governed by a quite general coupled mode formalism, we believe that the investigated Fano resonances can be realized using a wide range of structures such as micro-disk or micro-ring resonators[32]. The Fano concept may also be extended to other applications such as photodetectors and electro-optic modulators, with the prospect of largely reducing the energy consumption and increasing the bandwidth of on-chip integrated photonic interconnects[36]. By incorporating a quantum-dot in the cavity, the Fano structure can also be applied for quantum devices such as few-photon switch or sensor[37].

## Methods

### Fabrication of the InP PhC structure

Devices were fabricated using a process described elsewhere[35]. The PhC structure with triangular lattice was first patterned into a 500 nm positive resist (11% ZEP520A) by electron-beam lithography and then transferred to a 200 nm PECVD $SiN_x$ hard mask by $CHF_3/O_2$ reactive-ion etching. The resist was removed and the pattern was transferred to a 340 nm InP layer by a cyclic $CH_4/H_2$ and $O_2$ reactive-ion etching. The $SiN_x$ layer was removed by hydrofluoric acid and the air-slab structure was finally formed with diluted $H_2SO_4/H_2O_2$ solution (80 $H_2O$ : 1 $H_2SO_4$ : 8 $H_2O_2$) to selectively wet-etch a 1 μm $In_{0.53}Ga_{0.47}As$ sacrificial layer located below the InP membrane. Structures were fabricated to have a range of lattice constants, $a$ = 443-450 nm, and hole radii, $R/a$ = 0.2-0.29.

### Theoretical fitting

In order to estimate $\lambda_0$ and $Q_t$ for asymmetric Fano shapes, we used the following approach. For the Fano structures with mirror symmetry, the PTE transmission at the resonance $t_B^2$ was first estimated from FDTD calculations by comparing the light transmission of a reference PhC waveguide with and without the PTE. Based on least-squares method, the measured spectrum (normalized so that its off-resonance transmission equals $t_B^2$) was fitted to the expression in Eq. (2) to obtain $\lambda_0$ and $Q_t$. For the Fano structures with broken symmetry, besides estimating $t_B^2$ from FDTD calculations, the decay ratio $\gamma_2 / \gamma_1$ was estimated using three-dimensional finite element based simulations by comparing the electromagnetic flux leaking into the in- and output port when exciting the corresponding eigenmode of the cavity. The values for $\lambda_0$, $Q_t$ were then extracted by fitting Eq. (1) to the measured spectrum using the estimated values for $\gamma_2 / \gamma_1$ and $t_B^2$. The phase response curves calculated using these parameters showed good agreements with the measurements, confirming the validity of our model.

### Phase spectra measurements

We measured the phase response by comparing the microwave phase shift of the modulated envelope of the signal before and after the structure[38]. CW light from a tunable laser source was modulated at 10 GHz through a dual drive Mach-Zehnder modulator (MZM) driven by a microwave signal from a network analyzer, so that the generated signal is single-sideband. The input signal was set to TE polarization using a polarizer and fed into the device. The output of the device under test was sent to an InGaAs photodetector with a bandwidth of 50 GHz converting the signal back to the electrical domain and measured by the network analyzer (scattering parameter $S_{21}$), through which the microwave phase shift $\Delta\phi_{NA}(\omega)$ was obtained. By sweeping the input light from shorter to longer wavelength across the resonance, we obtain the group delay of the system $\tau_g(\omega) = -\partial\phi(\omega)/\partial\omega = \Delta\phi_{NA}(\omega)/(2\pi f_m)$, where $f_m$ = 10 GHz. Then $\phi(\omega)$ can be obtained simply by integration. In order to remove the phase shift contribution of the setup itself (i.e., fibers, couplers, amplifiers, etc.), the linear part of the phase response (delay) was estimated by characterizing a PhC reference waveguide (without cavity and PTE), resulting in the reference characteristic, $\phi_r(\omega)$. The phase shift originating from the cavity resonance was then obtained as $\phi_c(\omega) = \phi(\omega) - \phi_r(\omega)$. During the measurements, a polarization controller was employed to optimize the power launched from the laser into the MZM, an erbium-doped optical fiber amplifier (EDFA) and a variable optical attenuator were included at the input of the structure to tune the input optical power level. The input power level was estimated by subtracting the coupling loss of ~8.5 dB from the power at the input fiber. In Fig. 3 we used PhC structures with the resonance being wavelength-shifted compared to those in Figs. 1 and 2 because of the limited EDFA amplification range.

**All-optical modulation**

The experimental setup was similar to that in Ref. 22. For the nonlinear dynamics measurements, a 5 GHz RZ pump signal with 6% duty cycle was chosen as a good compromise between the need to obtain a sufficiently high peak power, for the given available average power of around 6 dBm at the input fiber, to get a sufficient modulation contrast, and the requirement that the pump spectrum does not overlap with the signal spectrum. Additionally, the relatively small repetition rate of 5 GHz ensures that carriers have recovered between subsequent pulses, simplifying the extraction of the signal recovery time. The CW signal and the modulated pump with their polarizations aligned to the TE mode were coupled in and out of the PhC devices through lensed single mode fibers. At the output of the device, the combined signal was first amplified by a standard in-line EDFA before the converted signal was selected with an optical band pass filter, detected with a photodiode with a bandwidth of 45 GHz and monitored with a sampling oscilloscope with a bandwidth of 70 GHz. The output signal waveform was averaged to obtain sufficient signal to noise ratio. The modulation contrast was estimated by using a noise floor given by the power level measured when the weak signal was off. For BER measurements, the pump signal was replaced by a 10 GHz RZ signal with 12% duty cycle, which was further modulated in the OOK format at 10 Gbit/s using a pulse pattern generator with $2^7-1$ or $2^{31}-1$ PRBS length before injection into the device.

**Acknowledgment**

We thank E. Palushani, D. Vukovic, J. Xu, J. Hua and P. Colman for assistance in establishing experimental set-ups, N. Kuznetsova for InP wafer preparations, and P. T. Kristensen and J. Rosenkrantz de Lasson for helpful discussions. The authors acknowledge financial support from Villum Fonden via the NATEC (NAnophotonics for TErabit Communications) Centre.

**Author Contributions**

Y. Y., M. H. and J. M. performed the theoretical analysis. M. H. and Y. Y. designed the device and performed the FDTD calculations. Y. Y. and K. Y. fabricated the samples. Y. Y., H. H., W. X., C. P. and L. K. O. performed the measurements. Y. C performed finite-element calculations. Y. Y., J. M. and M. H. prepared the manuscript. All authors commented on the manuscript. J. M. and K. Y. led the project.